\documentclass[aps,prl,showpacs,twocolumn,amsmath,amssymb,amsthm]{revtex4-1}
\usepackage{graphicx,bbm}

\newcommand{\be}{\begin{equation}}
\newcommand{\ee}{\end{equation}}

\newcommand{\bra}[1]{{\langle #1 \vert}}

\newcommand{\ket}[1]{{\vert #1 \rangle}}

\newcommand{\ii}{ {\rm i} }
\newcommand{\dd}{ {\rm d} }

\newcommand{\RR}{\mathbb{R}}
\newcommand{\CC}{\mathbb{C}}

\newcommand{\LL}{{\hat {\cal L}}}

\newcommand{\PP}{{\hat {\cal P}}}

\def\tr{{\,{\rm tr}\,}}
\def\Tr{{\,{\rm Tr}\,}}
\def\ad{{{\rm ad}\,}}
\def\one{\mathbbm{1}}

\def\PP{\hat{\cal P}}
\def\Pm{\mathbb{P}}

\def\Tm{\mathbb{T}}
\def\LL{\hat{\cal L}}
\def\DD{\hat{\cal D}}
\def\Hc{{\cal H}}
\def\Bc{{\cal B}}

\begin{document}

\title{$\Pm \Tm$-symmetric quantum Liouvillian dynamics}
\author{Toma\v{z} Prosen}
\affiliation{Department of Physics, FMF,  University of Ljubljana, Jadranska 19, 1000 Ljubljana, Slovenia}
\date{\today}

\begin{abstract}
We discuss a combination of unitary and anti-unitary symmetry of quantum Liouvillian dynamics, in the context of open quantum systems, which implies a $D_2$ symmetry of the complex Liouvillian spectrum. For sufficiently weak system-bath coupling it implies a uniform decay rate for {\em all} coherences, i.e., off-diagonal elements of the system's density matrix taken in the eigenbasis of the Hamiltonian.
As an example we discuss symmetrically boundary driven open $XXZ$ spin 1/2 chains.
\end{abstract}

 
\maketitle
 
{\em Introduction.--} The so-called PT symmetry, a product of a unitary and an anti-unitary transformation both of which square to identity, of non-Hermitian Hamiltonians has been introduced by Carl Bender \cite{bender} to study the spectral theory of Schr\" odinger-like operators and non-standard formulations of quantum mechanics. The generic possibilities \cite{bender,znojil,fleischmann,ali,west,schomerus} of having non-Hermitian operators with purely real spectra have recently found impressive experimental applications in non-linear optics \cite{christo,christo2} and even in LRC electric circuits \cite{lrc} where PT symmetry is achieved by a subtle combination of active elements, with the gain and the loss distributed in a symmetric way.

In this Letter we show that the concept of PT symmetry can be introduced, in contrast to Ref.~\cite{bender}, in the context of standard, orthodox quantum mechanics when one considers 
open, dissipative systems. We discuss a general situation in which a Liouvillian superoperator possesses a combination of unitary and anti-unitary master-symmetries 
(transformations in the linear space of generators of hermiticity preserving dynamical semi-groups). 
We show that, as a consequence of master symmetry, the spectrum of decay rates should have a dihedral ($D_2$) symmetry in the complex plane. Furthermore, for sufficiently weak system-bath coupling the spectrum has the shape of a cross, with one leg on the real axis, corresponding to dynamics of populations (diagonal operators) in the energy eigenbasis, and the other leg parallel to the imaginary axis, corresponding to the decay of coherences (off-diagonal operators), remarkably, all with the same asymptotic rate. This phenomenon offers a fundamentally new way of controlling decoherence as one deals with a single damping factor. Therefore, our general result should be of interest for a variety of fields which use the methods of open quantum systems \cite{breuer}, ranging from non-equilibrium statistical mechanics, quantum optics, quantum information, and quantum measurement theory, to condensed matter, high-energy theory and quantum cosmology.

As an example, we demonstrate how our symmetry is realized in the symmetrically boundary driven $XXZ$ spin 1/2 chain \cite{giuliano1,giuliano2,marko,prosen} which is described in terms of the standard Lindblad equation (with Hermitian Hamiltonian) and the canonical formalism of Markovian open quantum systems. We stress that our considerations assume only local-and-homogeneous-in-time open quantum system's theory, without any need for non-Hermitian central system's Hamiltonians.
Yet, the results we find formally faithfully generalize the mathematical framework of PT-symmetric quantum mechanics \cite{bender} to a Liouvillian setting.

{\em Quantum dynamics and Liouville space formalism.--} 
Let us consider a system defined on a {\em finite} Hilbert space $\Hc$ of $N$ states, with a canonical orthonormal (ON) basis $\ket{j},j=1\ldots N$. Let $\Bc(\Hc)$ denote the vector space of linear operators over $\Hc$. $N^2$ basis states of $\Bc(\Hc)$ shall be denoted by $E_{j,k} = \ket{j}\bra{k}$, $j,k=1\ldots N$.
Introducing the Hilbert-Schmidt inner product (see e.g. \cite{petz}) 
$ (\sigma,\rho) := \tr (\sigma^\dagger \rho)$,
$\Bc(\Hc)$ becomes a Hilbert space, with $\{ E_{j,k} \} $ being its ON basis. As arbitrary physical states are elements of $\Bc(\Hc)$ in the sense of density operators, the generator $\LL$ of quantum Liouvillian dynamics 
\begin{equation}
\frac{\dd}{\dd t}\rho(t) = \LL \rho(t)
\end{equation}  
can be considered as an element of $\Bc(\Bc(\Hc))$ and can be, say in a basis $\{E_{j,k}\}$, represented by $N^2\times N^2$ matrix.
For general Markovian open quantum systems, the Liouvillian $\LL$ can be always cast to the Lindblad form \cite{lindblad,breuer,zoller}
\begin{eqnarray}
\LL &=& -\ii (\ad H) + \gamma \DD, \quad (\ad H)\rho := [H,\rho], \label{eq:lindblad} \nonumber \\
\DD\rho &:=& \sum_m 2 L_m \rho L^\dagger_m - L^\dagger_m L_m \rho - \rho L^\dagger_m L_m, \label{eq:lind}
\end{eqnarray}
separating the unitary part generated by a {\em Hermitian} Hamiltonian $H\in\Bc(\Hc)$, and the dissipator $\DD\in \Bc(\Bc(\Hc))$, where $\{ L_m; m=1\ldots M\} \subset \Bc(\Hc)$ is a set of $M \le N^2-1$ Lindblad operators, which together with the system-bath {\em coupling strength}  $\gamma > 0$ contain all information that is left about the reservoirs (i.e., environment degrees of freedom) and coupling to them.
One can always adjust $\gamma$ such as to fix the trace of the dissipator $\Tr\DD = \sum_{j,k} (E_{j,k},\DD E_{j,k}) = -\Tr\hat{1}=-N^2$. 
We note that the abstract part of our discussion will not require a particular Lindblad form (\ref{eq:lindblad}) of the Liouvillian $\LL$, but we require only the existence of the following concepts.

{\em Liouvillian $\Pm$-transformation.--} Let $\PP \in \Bc(\Bc(\Hc))$ be some {\em unitary} transformation, $(\PP \sigma,\PP \rho) = (\sigma,\rho)$, $\forall \sigma,\rho\in\Bc(\Hc)$, squaring to identity $\PP^2 = \hat{1}$. Noting that $\PP^{-1} = \PP$, we define $\Pm$-transformation of the Liouvillian as
\begin{equation}
\Pm \LL := \PP \LL \PP.
\end{equation}

{\em Liouvillian $\Tm$-transformation.--} The Hilbert-Schmidt inner product completely defines the Hermitian adjoint of the Liouvllean $\LL^\dagger$, namely,
$(\sigma,\LL^\dagger \rho) = (\LL \sigma,\rho) = \overline{(\rho,\LL\sigma)}$, $\forall \sigma,\rho\in\Bc(\Hc)$. We define the $\Tm$-transformation as
\begin{equation}
\Tm \LL := \LL^\dagger.
\end{equation}
Note that the maps $\Pm$ and $\Tm$ can be considered as elements of $\Bc(\Bc(\Bc(\Hc)))$. By introducing the super-Hilbert-Schmidt inner product in $\Bc(\Bc(\Hc))$, $(\!(\hat{\cal X},\hat{\cal Y})\!) := \Tr(\hat{\cal X}^\dagger \hat{\cal Y})$, it can be verified straightforwardly that the map $\Pm$ is unitary, while the map $\Tm$ is anti-unitary, namely,
\begin{eqnarray}
(\!(\Pm\hat{\cal X},\Pm\hat{\cal Y})\!) &=& (\!(\hat{\cal X},\hat{\cal Y})\!),\\
(\!(\Tm\hat{\cal X},\Tm\hat{\cal Y})\!) &=& (\!(\hat{\cal Y},\hat{\cal X})\!), \qquad \forall \hat{\cal X},\hat{\cal Y}\in\Bc(\Bc(\Hc)).
\end{eqnarray}
{\em $\Pm\Tm$-symmetric Liouvillian.--}
Let $\LL'$ denote the {\em traceless} part of the Liouvillian, defined as
\begin{equation}
\LL = \LL' - \gamma \hat{1},\quad\textrm{with}\quad\gamma:=-\Tr\LL/\Tr\hat{1}.
\end{equation}
We shall define quantum Liouvillian dynamics as $\Pm\Tm$-symmetric if the following key identity holds
\begin{equation} 
\Pm\Tm\LL' = -\LL',
\label{eq:PTsym}
\end{equation}
or equivalently, $(\LL')^\dagger = -\PP \LL' \PP$. This $\Pm\Tm$ symmetry can be immediately applied to invert the Liouvillian propagator of generally dissipative dynamics
$\hat{\cal U}(t):=\exp(t\LL)$
\begin{equation}
\hat{\cal U}(-t) = e^{2\gamma t}\, \PP\, [\hat{\cal U}(t)]^\dagger\, \PP.
\end{equation}

Now we are in position to show the following result:\\
\noindent
{\bf Theorem.} {\em The spectrum of $\Pm\Tm$-symmetric Liouvillian (\ref{eq:PTsym}) has a dihedral group ($D_2$) symmetry in the complex plane, with the lines of symmetry $\ell_{\rm v} = -\gamma+ \ii \RR$ and $\ell_{\rm h}=\RR$. 
More specifically, writing the Liouvillian spectral decompositions
\begin{equation}
\LL  u_\alpha = \lambda_\alpha u_\alpha,\qquad
\LL^\dagger v_\alpha = \overline{\lambda}_\alpha v_\alpha
\label{eq:spcdec}
\end{equation}
with right and left eigenvectors which can be chosen bi-orthonormal, $(u_\alpha,v_\beta) = \delta_{\alpha,\beta}$, we state that for each eigenvalue of $\LL'$, $\lambda'_\alpha \equiv \lambda_\alpha + \gamma$, there exist the eigenvalues
$\lambda'_\beta =-\overline{\lambda'}_\alpha$ (image across $\ell_{\rm v}$), and $\lambda'_\eta = \overline{\lambda'}_\alpha$ (image across 
$\ell_{\rm h}$) with the right and left eigenvectors related via}
\begin{eqnarray}
u_\beta &=& \PP v_\alpha, \quad\, v_\beta = \PP u_\alpha, \label{eq:vsym} \\
u_\eta &=& u^\dagger_\alpha,\qquad v_\eta = v^\dagger_\alpha. \label{eq:hsym}
\end{eqnarray}
\noindent
In the case of a degenerate eigenvalue $\lambda_\alpha$, the eigenvectors $u_{\beta/\eta}$ and $v_{\beta/\eta}$ are to be understood as appropriate members of the right and left eigenspaces, respectively.

\noindent {\bf Proof.} The $\ell_{\rm v}$-reflection spectral symmetry with (\ref{eq:vsym}) is a direct consequence of $\Pm\Tm$ symmetry (\ref{eq:PTsym}), after applying it to the left-hand-sides of Eqs.~(\ref{eq:spcdec}), and multiplying the resulting equation by $\PP$.
The $\ell_{\rm h}$-reflection symmetry with (\ref{eq:hsym}), on the other hand, is an immediate consequence of hermiticity preservation of Liouvillian quantum dynamics,
$\LL(\rho^\dagger) = (\LL \rho)^\dagger$, which  clearly holds for the Lindbladian (\ref{eq:lindblad}), but also for any other (possibly non-Markovian) meaningful quantum dynamics \cite{zoller}. $\Box$

{\em Remarks.--}
One of the most interesting objects of open quantum dynamics is the steady state, $\rho(t\to\infty) = u_1$ corresponding to eigenvalue $\lambda_1 = 0$, which always exists, due to trace preservation in ${\cal H}$, which can be expressed as $\LL^\dagger v_1 = 0$ with $v_1 = \one = \sum_{j=1}^N \ket{j}\bra{j}$. This means that Eq. (\ref{eq:vsym}) yields already a non-trivial result, namely that 
 the fastest decaying mode $\lambda_{N^2} = -2\gamma$ is $u_{N^2} = \PP(\one)$.

Even more remarkable observation is that for sufficiently small coupling $\gamma$, say below some critical value $\gamma < \gamma_{\rm PT}$, the spectrum of decay modes lies strictly on the {\em cross}, $\{ \lambda_\alpha\} \subset \ell_{\rm v} \cup \ell_{\rm h}$. 
First, we show that if an eigenvalue 
$\lambda_\alpha(\gamma)$ lies on the real line $\ell_{\rm h}$, it remains on $\ell_{\rm h}$ as long as the eigenvalue is {\em isolated}. First order non-degenerate perturbation theory tells us that $\dd \lambda_{\alpha}/\dd \gamma = (v_\alpha, \DD u_\alpha)$, whence hermiticity conservation $\DD(u_\alpha)^\dagger = \DD(u^\dagger_\alpha)$, and $u_\alpha^\dagger = u_\alpha$, $v_\alpha^\dagger=v_\alpha$ following from (\ref{eq:hsym}) since $\overline{\lambda}_\alpha=\lambda_\alpha$, implies 
$\dd \overline{\lambda}_{\alpha}/\dd \gamma = \dd \lambda_{\alpha}/\dd \gamma \in\RR$. 
Second, we show similarly that for an isolated eigenvalue $\lambda_\alpha(\gamma)$ initially on $\ell_{\rm v}$, i.e., $\lambda'_\alpha \in \ii\RR$, we have $\dd\lambda'_\alpha/\dd\gamma \in \ii \RR$ as a simple 
consequence of the $\Pm\Tm$ symmetry of the traceless part of the dissipator $\DD'=\DD + \hat{1}$, $\Pm\Tm\DD' = -\DD'$. 
Namely, from $\overline{\lambda'}_\alpha=-\lambda'_\alpha$ and (\ref{eq:vsym}) follows that $u_\alpha=\PP v_\alpha$, so $\dd \overline{\lambda'}_\alpha/\dd\gamma = (\DD' u_\alpha,v_\alpha) = -(u_\alpha,\PP \DD' \PP v_\alpha) = -(v_\alpha,\DD' u_\alpha) = -\dd\lambda'_\alpha/\dd\gamma$.
The spectrum can then leave the cross  $\ell_{\rm v} \cup \ell_{\rm h}$, when at some $\gamma=\gamma_{\rm PT}$ a pair of eigenvalues collides and shoots off into the complex plane. 
By observing just the motion of the spectral points on the vertical leg $\ell_{\rm v}$ (while the argument should be quite similar for $\ell_{\rm h}$) the critical coupling strength where the $\Pm\Tm$ symmetry of the spectrum is spontaneously broken can be estimated heuristically as
\begin{equation}
\gamma_{\rm PT} \sim \|\DD'\|^{-1} d^{-2}.
\label{eq:heuristic}
\end{equation}
Here  $\|\DD'\|$ is the operator norm of the dissipator which estimates the maximal velocities $|\dd\lambda_\alpha/\dd\gamma|$ and $d$ denotes a typical density of states of $H$, $d^2$ giving a typical density of energy differences $\epsilon_j-\epsilon_k$.

What remains to be shown to prove this picture is that initially, for (infinitesimally) small $\gamma$, the eigenvalues $\lambda'_\alpha$ indeed start on $\RR \cup \ii \RR$. This is again easy to demonstrate, working in the eigenbasis of the Hamiltonian $H$, $H\ket{\psi_j}=\epsilon_j\ket{\psi_j}$, $j=1\ldots N$.
For $\gamma=0$, $\lambda'_\alpha = \ii (\epsilon_j - \epsilon_k)$ where $\alpha$ now labels all $N^2$ pairs of $j,k$. Let us assume that the energy spectrum $\epsilon_j$ is non-degenerate, i.e., all $\epsilon_j$ are different. Then we have exactly $N$ diagonal eigen-operators 
$d_j = \ket{\psi_j}\bra{\psi_j} \equiv u_\alpha = v_\alpha $ for which $\lambda'_\alpha = 0$. In order to understand their motion as we switch on $\gamma$, we have to solve first-order-degenerate-perturbation problem; i.e., we have to diagonalize an $N\times N$ matrix $V_{j,k}:=(d_j,\DD' d_k)$, $V a_\alpha = \xi_\alpha a_\alpha$, where each eigenvalue $\xi_\alpha$ determines the motion of some $\lambda'_\alpha$, $\dd \lambda'_\alpha/\dd \gamma|_{\gamma=0} = \xi_\alpha$, and the corresponding eigenvector determines the hybridization, $u_\alpha=v_\alpha = \sum a_{\alpha,j} d_j$. The key observation now is that the matrix $V$ is {\em real} and {\em symmetric} \cite{note}
\begin{eqnarray}
&&V_{j,k} = (d_j,\DD' d_k) = \bra{\psi_j}\DD'(\ket{\psi_k}\bra{\psi_k})\ket{\psi_j} \nonumber \\
&&=\!\sum_m \Bigl( 2 |\bra{\psi_j}L_m\ket{\psi_k}|^2\!-\!\bra{\psi_j}L^\dagger_m L_m\ket{\psi_j} \!-\! \bra{\psi_k}L^\dagger_m L_m\ket{\psi_k} \Bigr) \nonumber \\
&& +\, \delta_{j,k} = \overline{V}_{j,k} = V_{k,j},
\label{eq:realsymmetric}
\end{eqnarray}
proving that these $N$ Liouvillian eigenvalues $\lambda'_\alpha(\gamma)$ remain on the real line. Asymptotically, for small $\gamma$ the states $u_\alpha(\gamma)$ 
are diagonal and correspond to {\em populations} in the energy eigenbasis. The other $N^2-N$ eigenvalues $\lambda'_\alpha(\gamma)$, for $\epsilon_j\neq\epsilon_k$, move on the imaginary line, as shown in the previous paragraph, provided that they are isolated initially for $\gamma=0$, i.e., provided that all the energy spacings $\epsilon_j-\epsilon_k$ are
different (non-degenerate). These eigenvalues, asymptotically for small $\gamma$, correspond to off-diagonal operators -- {\em coherences} -- in the energy eigenbasis, and {\em all} decay with exactly the same rate ${\rm Re} \lambda_\alpha \equiv -\gamma$.
This is remarkable and should have experimentally observable consequences, e.g. one should be able to control the decoherence in such a system by handling a single damping factor $e^{-\gamma t}$. 

The above scenario which predicts the full Liouvillian spectrum of decay modes to belong to the cross $\ell_{\rm v} \cup \ell_{\rm h}$ for some non-empty coupling strength interval $0 \le \gamma \le \gamma_{\rm PT}$ is strictly justified only if the two conditions are met, namely that both, the energy spectrum $\{\epsilon_j\}$ and the energy difference spectrum $\{\epsilon_j-\epsilon_k; j\neq k\}$ are non-degenerate.
In the non-generic case when we have a degeneracy in either of the two, it can happen (in the absence of additional selection rules) that already infinitesimal dissipation $\gamma$ moves the corresponding  Liouvillian eigenvalues out into the complex plane, so the spontaneous $\Pm\Tm$ symmetry breaking of the Liouvillian spectrum may then occur already for a vanishing perturbation $\gamma_{\rm PT}=0$.

\begin{figure}
          \centering	
	\includegraphics[width=0.85\columnwidth]{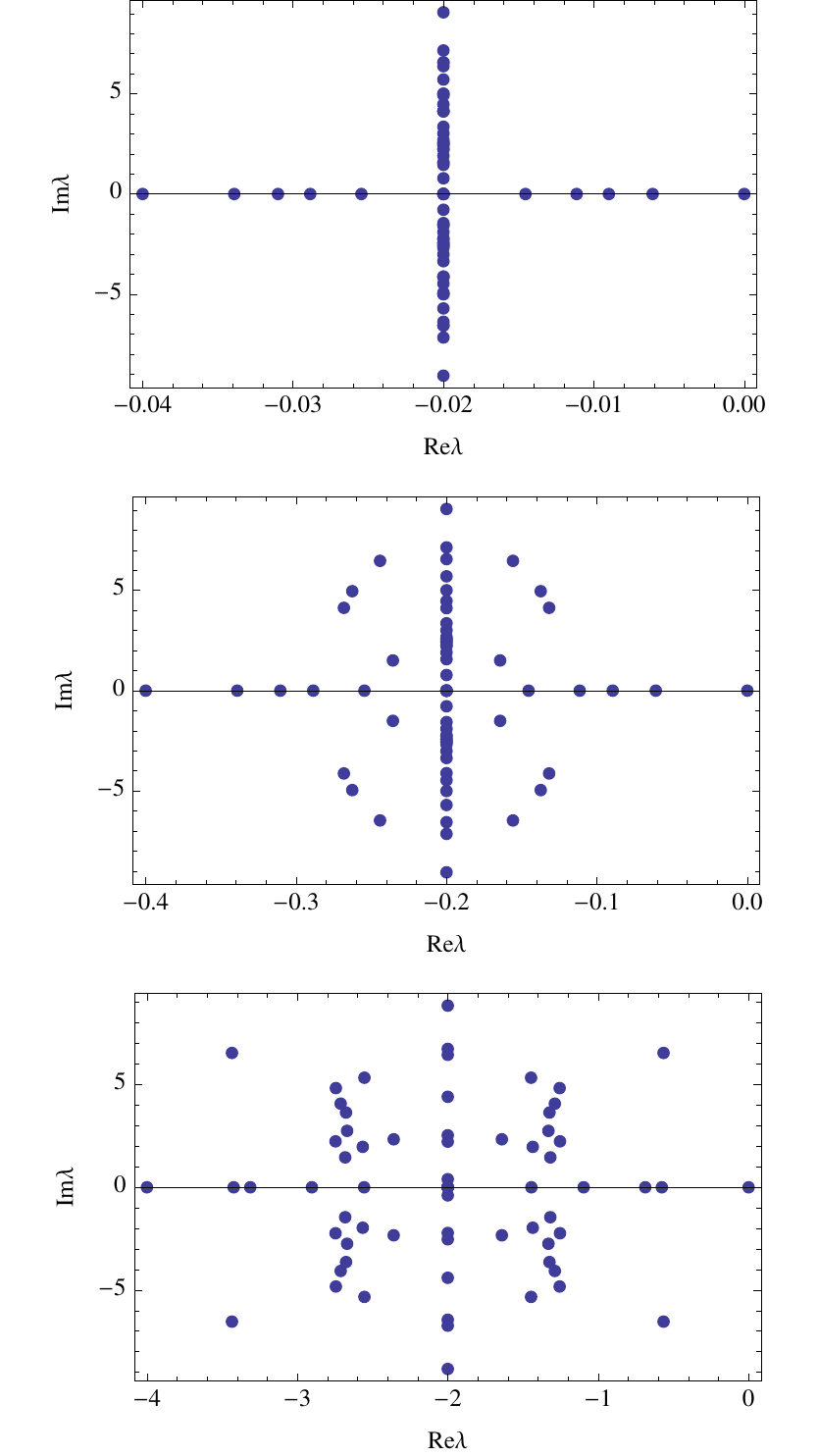}
	\caption{The Liouvillian spectrum $\{\lambda_k\}$ for an open $XXZ$ chain with $\Delta=1/2$, $n=4$, maximum driving $\mu=1$, 
	and $\gamma=0.02 < \gamma_{\rm PT}$ (top), 
	$\gamma=0.2 > \gamma_{\rm PT}$ (middle), and $\gamma=2 > \gamma_{\rm PT}$ (bottom). Note the dihedral symmetry of the spectrum with the horizontal 
	and vertical symmetry 
	lines, $\ell_{\rm h}=\RR$ and $\ell_{\rm v}=-\gamma+\ii\RR$, respectively.
	Because the $XXZ$ chain has a global conservation law $M^{\rm z}=\sum_j \sigma^{\rm z}_j$, we consider only the most relevant sector with zero total magnetization 
	$M^{\rm z}=0$.
	}
	\label{fig:spc}
\end{figure}

{\em Example.--}
We close by demonstrating our constructions in an interesting example, namely we consider an open $XXZ$ chain of $n$ spins 1/2 with the Hamiltonian
\be
H = \sum_{j=1}^{n-1} (2\sigma^+_j \sigma^-_{j+1} + 2\sigma^-_j \sigma^+_{j+1} + \Delta \sigma^{\rm z}_j \sigma^{\rm z}_{j+1})
\ee
where $\sigma^\pm_j = \frac{1}{2}(\sigma^{\rm x}_j\pm \ii \sigma^{\rm y}_j),\sigma^{\rm z}_j,j=1\ldots n$ are Pauli operators on a product space
${\cal H}=(\CC^2)^{\otimes n}$, with symmetric Lindblad driving acting on the edges of the chain only
\be
L_{1,2} = \frac{1}{2}\sqrt{1\pm\mu}\, \sigma^\pm_1,\quad
L_{3,4} = \frac{1}{2}\sqrt{1\mp\mu}\, \sigma^\pm_n,
\ee
with $M=4$, where $\mu\in [-1,1]$ is a {\em driving parameter} determining the magnetization bias between the left and the right bath. We have here $N=2^n$. This model has been intensively studied recently \cite{giuliano1,giuliano2,marko,prosen,popkov}, and it has been shown to admit exact solutions \cite{prosen} in the limiting cases of small $\gamma$ or $\mu=1$, and can exhibit diffusive spin transport (for $|\Delta|>1$) in the linear response regime \cite{marko}, of small $\mu$ and non-small $\gamma\sim 1$.
In order to disclose explicitly the $\Pm\Tm$ symmetry of such a symmetrically boundary driven $XXZ$ chain, it is instructive to identify the operator space 
${\cal B}(\cal H)$ with the tensor product ${\cal H}\otimes{\cal H}$, via the isomorphism 
\begin{equation}
\ket{\psi}\bra{\phi} \leftrightarrow \ket{\psi} \otimes S \ket{\phi}
\end{equation} 
where
$S:=\prod_{j=1}^n \sigma^{\rm x}_j$ is a global spin-flip operation in $\sigma^{\rm z}_j$ eigenbasis. Then the Liouvillian is represented as
\begin{eqnarray}
\LL &\leftrightarrow&  \one \otimes \left(\ii H -  \frac{\gamma\mu}{4}(\sigma^{\rm z}_1\!-\!\sigma^{\rm z}_n)\!\right) - \left(\ii H -  \frac{\gamma\mu}{4}(\sigma^{\rm z}_1\!-\!\sigma^{\rm z}_n)\!\right) \otimes \one
 \nonumber \\
&& + \frac{\gamma(1+\mu)}{2}(\sigma^+_1 \otimes \sigma^-_1 + \sigma^-_n \otimes \sigma^+_n) \nonumber \\
&& + \frac{\gamma(1-\mu)}{2}(\sigma^-_1 \otimes \sigma^+_1 + \sigma^+_n \otimes \sigma^-_n) - \gamma \one\otimes \one.
\label{eq:exprL}
\end{eqnarray}
Let us define the parity transformation $\PP$, such that it corresponds to the following operator
\begin{equation}
\PP \leftrightarrow \biggl(R \prod_{j=1}^n \sigma^{\rm_z}_j\biggr)\otimes R
\end{equation}
where $R$ is a reflection permutation which reverses the order of sites $j \longleftrightarrow n+1-j$, i.e., in the common eigenbasis of $\sigma^{\rm z}_j$, $j=1\ldots n$, it reads
$R = \sum_{m_1\ldots m_n \in\{+1,-1\}} \ket{m_n\ldots m_1}\bra{m_1\ldots m_n}$.
In notation of note \cite{note}: $U \equiv R\prod_j \sigma^{\rm z}_j$, $W\equiv R$.
It takes a straightforward calculation to show that indeed $\PP^2=\hat{1} \leftrightarrow \one \otimes \one$, and $(\LL+\gamma\hat{1})^\dagger = -\PP (\LL+\gamma\hat{1})\PP$. In fact, the $\Pm \Tm$ symmetry (\ref{eq:PTsym}) holds for each of the three rows of the expression (\ref{eq:exprL}) separately.
In Fig.~\ref{fig:spc} we show three different Liouvillian spectra for different values of the coupling constant, before and after the transition $\gamma=\gamma_{\rm PT}$. Numerical experiments indicate that the critical value in the leading order decays exponentially with the chain length, $\gamma_{\rm PT} \propto d^{-2} \propto 4^{-n}$,  as estimated in (\ref{eq:heuristic}). We note that even for $\gamma \gg \gamma_{\rm PT}$, a substantial fraction of spectral points remain on the line $\ell_{\rm v} = -\gamma + \ii\RR$, hence a significant spectral weight for uniform relaxation $e^{-\gamma t}$ is expected for typical (offdiagonal) observables. Remarkably, an important operator in the transport theory, the spin current operator $J=\ii \sum_{j=1}^{n-1} (\sigma^+_j \sigma^-_{j+1}- \sigma^-_j \sigma^+_{j+1})$,
has vanishing diagonal matrix elements in the energy eigenbasis, so its expectation value relaxes with a uniform rate $|\tr [(\rho(t)-\rho(\infty)) J]| \sim e^{-\gamma t}$, for $\gamma < \gamma_{\rm PT}$.

{\em Discussion.--} We outlined a general framework for analysis of a combined unitary ($\Pm$) and anti-unitary ($\Tm$) master-symmetry
of the most general types of quantum master equations which are local in time. We stress that our analysis remains strictly in the framework of canonical quantum mechanics, so we need no active elements or non-Hermitian system's Hamiltonians for our constructions. $\Pm\Tm$ symmetry of dissipative Liouvillian dynamics can thus occur only with respect to a shift parallel to the imaginary line which represents an average damping rate. In the asymptotic regime of weak system-bath coupling, the Lioivillean spectrum can be strictly separated with the coherences -- the off-diagonal matrix elements of the state  in the energy eigenbasis -- decaying with a strictly uniform rate. We discussed a simple explicit example of $\Pm\Tm$-symmetric Liouvillian dynamics in open $XXZ$ spin chains, but other interesting and experimentally accessible realizations are possible. For example, the recently studied symmetrically driven fermi Hubbard chain \cite{giuliano2,hub} is $\Pm\Tm-$symmetric as well (as can be easily seen in spin-ladder formulation), but applications to driven open bosonic cold atom systems should also be possible.

Inspiring discussions with Tsampikos Kottos, which initiated this work, and with Corinna Kollath and Thomas Seligman are warmly acknowledged. Research has been sponsored by the grants P1-0044 and J1-2208 of the Slovenian Research Agency (ARRS).


\begin{thebibliography}{10}

\bibitem{bender} C.~M.~Bender and S.~Boettcher, Phys. Rev. Lett. {\bf 80}, 5243 (1998);
C.~M.~Bender, Rep. Prog. Phys. {\bf 70}, 947 (2007).

\bibitem{znojil} M.~Znojil, Phys. Lett. A {\bf 259}, 220 (1999); G.~Leval and M. Znojil, J. Phys. A: Math. Gen. {\bf 33}, 7165 (2000).

\bibitem{fleischmann} O.~Bendix, R.~Fleischmann, T.~Kottos, and B.~Shapiro, Phys. Rev. Lett. {\bf 103}, 030402 (2009).

\bibitem{ali} A.~Mostafazadeh, Phys. Rev. Lett. {\bf 102}, 220402 (2009).

\bibitem{west} C.~T.~West, T.~Kottos, and T. Prosen, Phys. Rev. Lett. {\bf 104}, 054102 (2010).

\bibitem{schomerus} H.~Schomerus, Phys.~Rev.~Lett.~{\bf 104}, 233601 (2010); Phys. Rev. A {\bf 83}, 030101(R) (2011).

\bibitem{christo} K.~G.~Makris, R.~El-Ganainy, D.~N.~Christodoulides and Z.~H.~Musslimani, Phys. Rev. Lett. {\bf 100}, 103904 (2008);
A.~Guo, {\em et al.}, Phys. Rev. Lett. {\bf 103}, 093902 (2009).

\bibitem{christo2} Z.~Lin, H.~Ramezani, T.~Eichelkraut, T.~Kottos, H.~Cao, and D.~N.~Christodoulides, Phys. Rev. Lett. {\bf 106}, 213901 (2011).

\bibitem{lrc} J.~Schindler, A.~Li, M.~C.~Zheng, F.~M.~Ellis, and T.~Kottos, Phys. Rev. A {\bf 84}, 040101 (2011);
Z.~Lin, J.~Schindler, F.~M.~Ellis, and T.~Kottos, {\em ibid} {\bf 85}, 050101 (2012);
H.~Ramezani, J.~Schindler, F.~M.~Ellis, U.~GŸnther, and T.~Kottos, {\em ibid} {\bf 85}, 062122 (2012).

\bibitem{breuer} H.-P. Breuer and F. Petruccione, {\em The theory of open quantum systems}, (Oxford University Press,  New York 2002).

\bibitem{giuliano1} G.~Benenti, G.~Casati, T.~Prosen, and D.~Rossini, Europhys. Lett. {bf 85}, 37001 (2009).

\bibitem{giuliano2} G.~Benenti, G.~Casati, T.~Prosen, D.~Rossini, and M.~\v Znidari\v c, Phys. Rev. B {\bf 80}, 035110 (2009).

\bibitem{marko} T. Prosen and M. \v Znidari\v c, J. Stat. Mech {\bf 2009}, P02035 (2009); M. \v Znidari\v c, Phys. Rev. Lett. {\bf 106}, 220601 (2011).

\bibitem{prosen} T. Prosen, Phys. Rev. Lett. {\bf 106}, 217206 (2011); {\em ibid} {\bf 107}, 137201 (2011).

\bibitem{petz} D. Petz, Linear Algebra Appl. {\bf 244}, 81 (1996).

\bibitem{lindblad} G.~Lindblad, Commun. Math. Phys. {\bf 48}, 119 (1976);
V.~Gorini, A.~Kossakowski, and E.~C.~G.~Sudarshan, J. Math. Phys. {\bf 17}, 821 (1976).

\bibitem{zoller} C.~W.~Gardiner and P.~Zoller, {\em Quantum Noise: A Handbook of Markovian and Non-Markovian Quantum Stochastic Methods with Applications to Quantum Optics},
(Springer-Verlag, Berlin Heidelberg 2004).

\bibitem{note} While $V_{j,k} \in \RR$ follows directly from (\ref{eq:realsymmetric}), a strict proof of symmetry $V_{j,k}=V_{k,j}$ requires an extra assumption, which also provides a useful hint for concrete realizations of $\Pm\Tm$ symmetry. Namely, we claim that the Liouvillian flow is $\Pm\Tm$-symmetric, if: (i) $\PP\rho = U \rho W$ where $U,W\in {\cal B}({\cal H})$ are two unitary operators satisfying $U^2=W^2=\one$, and $[H,U]=[H,W]=[U,W]=0$, (ii) $\exists$ a self-inverse one-to-one map $\pi : \{1\ldots M\} \to \{1\ldots M\}$, $\pi(\pi(m))\equiv m$, such that $U L_m = -\nu_m L_{\pi(m)}^\dagger U$ and $W L_m = \nu_m L_{\pi(m)}^\dagger W$, with $\nu_m\in\{\pm 1\}$, and (iii) $\{L_m,L_m^\dagger\} = c_m \one$, $c_m \in \RR$.
Namely, using (iii), $\DD'$ can be written as $\DD'\rho = \sum_m (2 L_m \rho L_m^\dagger - \frac{1}{2}\{ [L^\dagger_m,L_m],\rho\})$, and using (ii), $\DD' \PP \rho = \DD'(U\rho W)=- U \sum_m (2 L_m^\dagger \rho L_m - \frac{1}{2}\{[L^\dagger_m,L_m],\rho\})W = -\PP (\DD')^\dagger \rho $. Finally, using (i), we have also $(\ii\,\ad H)\PP = \PP(\ii\,\ad H) = -\PP (\ii\,\ad H)^\dagger$, and therefore $\LL'\PP = -\PP(\LL')^\dagger$, i.e., (\ref{eq:PTsym}).

Now, in order to prove $V_{j,k}=V_{k,j}$ in Eq.~(\ref{eq:realsymmetric}), we only need to show $\sum_m |\bra{\psi_j}  L_m \ket{\psi_k}|^2 = \sum_m |\bra{\psi_k}  L_m \ket{\psi_j}|^2$. Indeed, due to (i)
 the eigenvectors $\ket{\psi_j}$ of $H$ can be chosen to be simultaneously eigenvectors of $W$, say $W \ket{\psi_j} = \omega_j \ket{\psi_j}$, with $\omega_j \in\{\pm 1\}$, 
so we have $|\bra{\psi_j}  L_m \ket{\psi_k}|^2 = |\bra{\psi_j} L_m W \ket{\psi_k}|^2 =  |\bra{\psi_j} W L_{\pi(m)}^\dagger \ket{\psi_k}|^2 = |\bra{\psi_k}L_{\pi(m)}\ket{\psi_j}|^2$.  $\Box$

\bibitem{popkov} V.~Popkov, M.~Salerno, and G.~M.~Sch\" utz, Phys. Rev. E {\bf 85}, 031137 (2012).

\bibitem{hub} T. Prosen and M. \v Znidari\v c, e-print {\tt arXiv:1203.1727}; Phys. Rev. B, {\em to appear}.

\end{thebibliography}
\end{document}